\documentclass[aps,prl,twocolumn,superscriptaddress]{revtex4-2}

\usepackage{amsmath,amssymb,amsfonts}
\usepackage{mathtools}
\usepackage{graphicx}
\usepackage{hyperref}
\usepackage{bbm}
\usepackage{amsthm}
\usepackage{booktabs}

\newcommand{\HH}{\mathbb{H}_+}

\newcommand{\NN}{\mathcal{N}}

\newcommand{\HeffSP}{H_{\mathrm{eff}}^{\mathrm{SP}}}
\newcommand{\HeffSYM}{H_{\mathrm{eff}}^{\mathrm{SYM}}}
\DeclareMathOperator{\Real}{Re}
\DeclareMathOperator{\Imag}{Im}

\begin{document}

\title{Dispersion Relations Across the Unitarity Boundary}

\author{Kejun Liu}
\email{kjliu@suda.edu.cn}
\affiliation{State Key Laboratory of Bioinspired Interface Material Science, Institute of Nano \& Functional Materials, Soochow University, Suzhou 215123, China}
\affiliation{School of Physical Science and Technology, Soochow University, Suzhou 215006, China}

\date{\today}

\begin{abstract}
Kramers--Kronig (KK) relations rest on a binary premise: a response function is either analytic in the upper half-plane or it is not. We show that a single reduced-state transform organizes both outcomes into a sharp dichotomy controlled by microscopic unitarity. One closed-form function carries, simultaneously, a zero and a pole in the upper half-plane; the spectral abscissa $\alpha$ of the reduced propagator decides which is realized. For $\alpha\le0$ (unitary reduction) the upper-half-plane object is a \emph{protected zero}: KK holds, yet the zero is directly measurable from a finite-time coherence record by a damped Fourier transform---no analytic continuation---and obeys a closed law $\Imag\zeta=0.3092\,g$. For $\alpha>0$ (gain-driven non-unitary reduction) the zero is replaced by a genuine pole, the Blaschke winding number jumps from $0$ to $1$, and KK acquires a Lorentzian residue correction scaling as $\Delta_{\rm KK}\sim|\gamma-\gamma_c|^{\nu}$, $\nu\approx-1.08$, peaking at threshold. The protected zero is not inert: any scalar single-channel kernel extraction (MKCT, Shi--Geva, process tensor) is forced to reproduce a \emph{phantom resonance}---a refractive feature with no absorptive origin, at a protocol-independent frequency $\Real\zeta$---without any initial system--bath correlation. We give the closed-form criteria, a measurable terahertz signature, and the exactly solvable dimer and Jaynes--Cummings models that realize both sides of the boundary.
\end{abstract}

\maketitle

% ============================================================
Causality in linear response---analyticity of the response function in the upper half-plane $\HH$---is one of the most universal constraints in physics~\cite{toll1956,nussenzveig1972}. It is conventionally a binary property: a response either obeys the upper-half-plane analyticity required for standard Kramers--Kronig (KK) relations, or it does not. In passive, unitary systems the boundedness of the time-evolution operator guarantees the former. In non-unitary reduced dynamics---gain media, driven-dissipative systems, non-Hermitian effective Hamiltonians---the guarantee can fail~\cite{ruter2010,peng2014ptmicro,ozdemir2019review,el-ganainy2018review}.

We show that the binary picture hides a sharp dichotomy. Let $\tilde\sigma(z)$ be the Laplace transform of a reduced propagator and let $\alpha=\limsup_{t\to\infty}t^{-1}\log\|\sigma(t)\|_{\rm op}$ be its spectral abscissa. A single closed-form reduced-state transform carries, at once, a zero and a pole in $\HH$; the value of $\alpha$ decides which one governs the physics. The two regimes---and the boundary between them---are the subject of this Letter:

\begin{itemize}
\item \textbf{Unitary side} ($\alpha\le0$): the upper-half-plane object is a \emph{zero} of $\tilde\sigma$. KK holds, but the zero is directly measurable and forces a phantom KK anomaly in any scalar reconstruction.
\item \textbf{Non-unitary side} ($\alpha>0$): a genuine \emph{pole} enters $\HH$, the Blaschke winding number jumps $0\to1$, and KK acquires a residue correction with a critical scaling exponent.
\item \textbf{Boundary} ($\alpha=0$): microscopic unitarity. The zero and the pole are the two faces of one function across this line.
\end{itemize}

This reorganizes phenomena previously studied in isolation---PT-symmetric pole migration~\cite{ruter2010,ozdemir2019review}, apparent acausality from reduced descriptions~\cite{zhang2024internal,gavassino2026}---as the two sides of a single, sharply bounded structure, and it makes the unitary side \emph{measurable}.

\begin{figure}[t]
\includegraphics[width=\columnwidth]{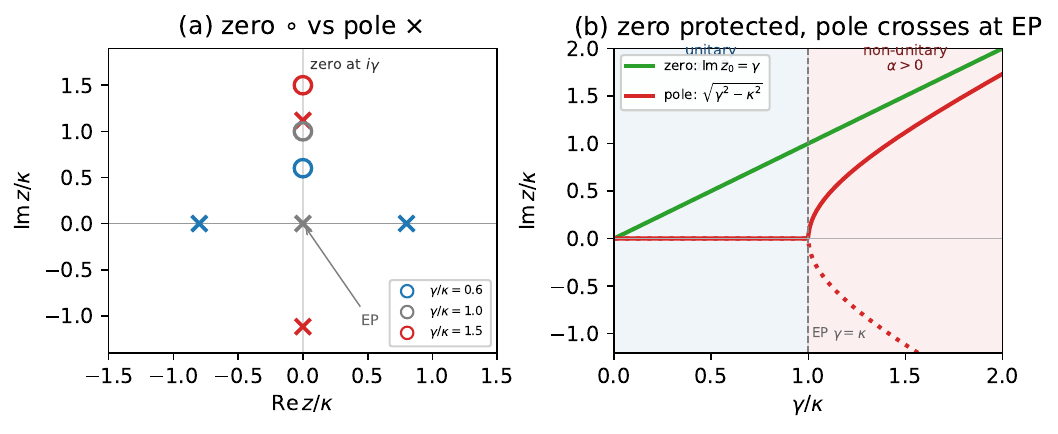}
\caption{The dichotomy carried by Eq.~\eqref{eq:sigma11}. As $\gamma/\kappa$ sweeps across the exceptional point, the always-present UHP zero $z_0=i\gamma$ (open circles) coexists with the pole pair $z_\pm$ (filled), which migrates from the real axis into $\HH$ at $\gamma=\kappa$. Color encodes $\gamma/\kappa$. The unitary side ($\alpha=0$, $\gamma<\kappa$) carries only the zero; the non-unitary side ($\alpha>0$) acquires the pole and $\NN_B:0\to1$.}
\label{fig:dichotomy}
\end{figure}

% ============================================================
{\em One function, two structures.}---Both sides descend from one object. Take the PT-symmetric dimer $H(\gamma)=\bigl(\begin{smallmatrix}-i\gamma&\kappa\\ \kappa&i\gamma\end{smallmatrix}\bigr)$, $[\mathcal{PT},H]=0$, and the reduced single-site amplitude $\sigma_{11}(t)=\langle1|e^{-iHt}|1\rangle$. Its Laplace transform is, in closed form,
\begin{equation}
\tilde\sigma_{11}(z)=\frac{i\,(z-i\gamma)}{z^2+\gamma^2-\kappa^2},
\label{eq:sigma11}
\end{equation}
which carries both structures at once:
\begin{itemize}
\item a \emph{zero} at $z_0=i\gamma\in\HH$ for all $\gamma>0$;
\item \emph{poles} at $z_\pm=\pm\sqrt{\kappa^2-\gamma^2}$, real for $\gamma<\kappa$ but in $\HH$ for $\gamma>\kappa$.
\end{itemize}
The eigenvalues of $H$ are $\pm\sqrt{\kappa^2-\gamma^2}$: real for $\gamma<\kappa$ (spectral abscissa $\alpha=0$, bounded evolution) and $\pm i\sqrt{\gamma^2-\kappa^2}$ for $\gamma>\kappa$ ($\alpha>0$, one growing mode). The poles of $\tilde\sigma$ are exactly the resolvent poles, whose supremal imaginary part is $\alpha$. Hence
\begin{equation}
\|\sigma(t)\|_{\rm op}\ \text{bounded}\iff\alpha\le0\iff\text{no UHP pole}\iff\NN_B=0,
\label{eq:dichotomy-chain}
\end{equation}
where $\NN_B=(2\pi i)^{-1}\oint(\tilde\sigma'/\tilde\sigma)\,dz$ is the Blaschke winding number counting UHP poles. Microscopic unitarity ($\alpha\le0$) is necessary and sufficient for the upper-half-plane object to be a zero rather than a pole. The exceptional point $\gamma=\kappa$ is the boundary; the pole emerges continuously from the real axis as $\Imag z_+=\sqrt{\gamma^2-\kappa^2}\to0^+$, so the protected (zero) phase and the charged (pole) phase are the two ends of one family (Fig.~\ref{fig:dichotomy}).

\begin{figure}[t]
\includegraphics[width=\columnwidth]{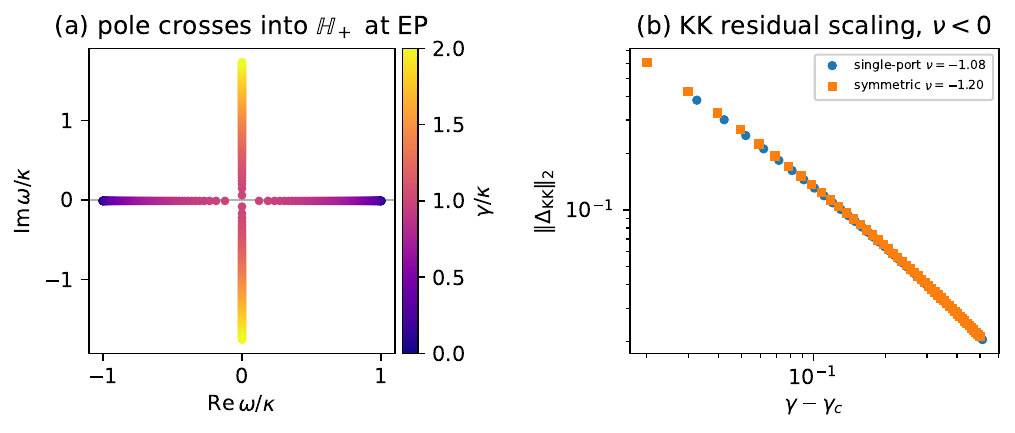}
\caption{Non-unitary side. (a)~Pole trajectory of $r(\omega;\gamma)$: a pole crosses from the LHP into $\HH$ at the EP. (b)~Log-log scaling of $\|\Delta_{\rm KK}\|_2$ vs $\gamma-\gamma_c$: single-port (circles) $\nu=-1.08$, symmetric (squares) $\nu=-1.20$. The violation peaks at threshold.}
\label{fig:nonunitary}
\end{figure}

% ============================================================
{\em Non-unitary side: pole migration and KK breakdown.}---On the gain side the dimer is coupled to a one-port waveguide. Integrating out the bus under the Markov approximation~\cite{gardiner1985,walls-milburn} gives $\HeffSP=H(\gamma)-i(\gamma_{\rm ex}/2)|1\rangle\!\langle1|$ and the exact reflection coefficient
\begin{equation}
r(\omega;\gamma)=1+i\gamma_{\rm ex}G_{11}^{\rm eff}(\omega),\quad G^{\rm eff}=(\omega I-\HeffSP)^{-1},
\label{eq:r-sp}
\end{equation}
with poles at the roots of $D^{\rm SP}(\omega)=\omega^2+i(\gamma_{\rm ex}/2)\omega+\gamma^2+\gamma\gamma_{\rm ex}/2-\kappa^2$. The pole enters $\HH$ when $\gamma(\gamma+\gamma_{\rm ex}/2)>\kappa^2$, and $\NN_B$ jumps from $0$ to $1$ (Fig.~\ref{fig:nonunitary}), protected by the codimension-one structure of the EP within PT-symmetric matrices~\cite{kawabata2019}. A response meromorphic in $\HH$ with simple poles $\{z_j\}$, residues $\{\rho_j\}$ admits the Blaschke factorization $r=B\cdot r_{\rm reg}$, $B(z)=\prod_j(z-z_j)/(z-z_j^*)$, and the residue-corrected KK relations
\begin{align}
\Real r(\omega)&=\tfrac{1}{\pi}\mathrm{P}\!\!\int\!\frac{\Imag r(\omega')}{\omega'-\omega}d\omega'+2\sum_j\Real\!\Big(\frac{\rho_j}{\omega-z_j}\Big),\\
\Imag r(\omega)&=-\tfrac{1}{\pi}\mathrm{P}\!\!\int\!\frac{\Real r(\omega')}{\omega'-\omega}d\omega'+2\sum_j\Imag\!\Big(\frac{\rho_j}{\omega-z_j}\Big).
\end{align}
The standard-KK residual $\Delta_{\rm KK}(\gamma)\equiv\|\Real r-\mathcal{H}[\Imag r]\|_2$ scales as
\begin{equation}
\Delta_{\rm KK}\sim|\gamma-\gamma_c|^{\nu},\quad\nu=-1.080\pm0.011,
\label{eq:nu}
\end{equation}
with $R^2>0.997$. The \emph{negative} exponent is the striking feature: as $\gamma$ rises above $\gamma_c$ the pole migrates deeper into $\HH$ and its residue shrinks, so the KK violation peaks at threshold and decays in the broken phase, correcting the heuristic $\nu=+1/2$. The residue correction reduces $\|\Delta_{\rm KK}\|_2$ by $\ge15\times$ at every $\gamma>\gamma_c$. (A four-site PT-SSH cross-check gives oscillatory scaling, $R^2=0.04$: the clean exponent is specific to the single-pole class~\cite{SM}.)

% ============================================================
{\em Unitary side: the protected zero is measurable.}---On the unitary side $\tilde\sigma$ stays in $\HH$ (Hardy) and its UHP feature is a zero, not a pole; standard KK holds for $\tilde\sigma$ itself~\cite{liu2026hardy}. The zero is nonetheless a physical observable. In a Jaynes--Cummings cavity-QED model $H_{\rm JC}=\tfrac{\omega_0}{2}\sigma_z+\omega_c a^\dagger a+g(\sigma_+a+\sigma_-a^\dagger)$, the coherence-channel transform $\tilde\sigma_{eg}(z)$ acquires a UHP zero $\zeta$ once the cavity carries Fock coherence~\cite{liu2026hardy}. Because, for $\eta=\Imag z>0$,
\begin{equation}
\tilde\sigma_{eg}(\xi+i\eta)=\int_0^\infty e^{i\xi t}\,e^{-\eta t}\,\sigma_{eg}(t)\,dt
\label{eq:damped-ft}
\end{equation}
is a low-pass-filtered Fourier transform of the measured coherence, the zero is recovered from a finite-time record on $[0,T]$ \emph{without} analytic continuation. The recovery converges fast: to $1\%$ at $T\approx20/\omega_c$ (about three Rabi periods) and to $10^{-6}$ at $T\approx80/\omega_c$, robust to signal-to-noise $\gtrsim300$. Its location obeys a closed, truncation-stable law,
\begin{equation}
\Imag\zeta=0.3092\,g,\qquad \Real\zeta=\omega_c\pm2.2041\,g,
\label{eq:closed-law}
\end{equation}
identical to five digits for cavity cutoffs $N_{\max}=2$--$5$: only the two lowest vacuum-Rabi manifolds set it. The fragile regime where $\Imag\zeta>0$ occupies the high-temperature, strong-coupling quadrant of the $(T,\lambda)$ plane~\cite{SM}. Observing a UHP zero certifies that the bath carries Fock coherence---a state-tomography diagnostic obtained without fitting any model, because a causal transform cannot have a UHP pole and a passive (Stieltjes) channel has all zeros real.

% ============================================================
{\em The phantom resonance.}---The protected zero is not inert. Any scalar single-channel force-fit that reproduces the measured coherence from a homogeneous Volterra equation has
\begin{equation}
\tilde K_{\rm eff}(z)=-iz-L_s-\frac{\sigma_{eg}(0)}{\tilde\sigma_{eg}(z)},
\label{eq:keff-scalar}
\end{equation}
so a UHP zero $\zeta$ of $\tilde\sigma_{eg}$ becomes a UHP \emph{pole} of $\tilde K_{\rm eff}$ with residue $R=-\sigma_{eg}(0)/\tilde\sigma_{eg}'(\zeta)$, and by the Plemelj formula the KK residual is the Lorentzian
\begin{equation}
\Delta(\omega)=2\,\Real\!\Big[\frac{R}{\omega-\zeta}\Big],
\label{eq:phantom}
\end{equation}
a ``phantom resonance'' centered at $\Real\zeta$---a frequency with \emph{no underlying transition}---of half-width $\Imag\zeta$. Three features make it a sharp prediction rather than a numerical accident.

\emph{One condition, not two.} The operator-valued NZ kernel $\tilde{\mathcal K}=\tilde{\mathcal K}+\tilde{\mathcal I}\tilde\sigma^{-1}$ acquires a pole only if the inhomogeneous term $\tilde{\mathcal I}\neq0$, i.e.\ only with initial system--bath correlation. The scalar one-channel kernel~\eqref{eq:keff-scalar} is a different object: reducing four channels to one routes the inter-channel coupling into the scalar kernel, so it acquires the phantom for \emph{any} factorized Fock-coherent state, with no correlation needed. Every standard extraction (MKCT, Shi--Geva, single-observable process tensor) performs this scalar fit and therefore sees the phantom.

\emph{A refractive feature with no absorptive origin.} Since $\tilde K_{\rm eff}$ has a UHP pole it is not a passive response function, and the KK residual~\eqref{eq:phantom} lives entirely on the dispersive ($\Real$) side. A spectroscopist who measures only absorption ($\Imag$) and reconstructs the refractive part via standard KK \emph{under-predicts} the measured $\Real$ by a Lorentzian at $\Real\zeta$: a refractive-index peak with no absorption peak that produced it---the operational signature that the fitted response is not passive.

\emph{Protocol independence.} The phantom sits at $\Real\zeta$, a property of the state transform $\tilde\sigma_{eg}$ alone, independent of how the kernel is extracted. MKCT, Shi--Geva, and process-tensor fits to the same $\sigma_{eg}(t)$ all reproduce it at the \emph{same} frequency: the phantom is a geometric marker, not a method artifact.

% ============================================================
{\em The boundary is sharp.}---Equation~\eqref{eq:dichotomy-chain} makes the unitary boundary exact: the upper-half-plane object is a zero iff $\alpha\le0$. The transition is not confined to the toy dimer. In the Jaynes--Cummings model with a true atomic gain channel $H_{\rm eff}=H_{\rm JC}+i\gamma_g|e\rangle\langle e|$, the same zero$\to$pole transition occurs in a genuine cavity-QED system: at $\gamma_g=0$ the coherence transform carries UHP zeros and no pole; for $\gamma_g>0$ genuine UHP poles appear with $\Imag(\text{pole})=\alpha=\gamma_g$~\cite{SM}. The dimer's threshold EP ($\alpha=\sqrt{\gamma^2-\kappa^2}$, square-root onset) and the JC gain channel ($\alpha=\gamma_g$, linear onset) realize the same boundary $\alpha=0$ through different approaches to it: the boundary is universal, the approach is model-dependent.

% ============================================================
{\em Measurable terahertz signature.}---The phantom is directly testable. For a vibrational-polariton or cavity-QED platform near vacuum Rabi splitting $\omega_{\rm Rabi}\sim2g$ in the THz band, Eq.~\eqref{eq:closed-law} fixes the phantom frequency and linewidth: a mid-coupling vibrational polariton ($\omega_c=15$~THz, $g/\omega_c=0.1$) gives a phantom of linewidth $\approx464$~GHz at $\Real\zeta\approx18.3$~THz; a THz cavity-QED system ($\omega_c=1$~THz) gives $\approx31$~GHz. The protocol: (i)~prepare a Fock-coherent cavity state (a thermal cavity is Fock-diagonal and carries no UHP zero); (ii)~measure the complex response by THz time-domain spectroscopy; (iii)~compute the standard-KK prediction from the measured absorption; (iv)~the residual is a Lorentzian at $\Real\zeta$ with no matching absorption peak---the phantom. On the non-unitary side, the same setup on a PT-symmetric metasurface~\cite{bhardwaj2025} or an electronic RLC dimer with a negative-impedance converter~\cite{schindler2011rlc} measures the pole migration through $\Delta_{\rm KK}\sim|\gamma-\gamma_c|^\nu$ with $\nu<0$.

% ============================================================
{\em Discussion.}---The dichotomy gives a common language for causality in open systems. Apparent acausality from reduced descriptions~\cite{zhang2024internal} or classical coarse-graining~\cite{gavassino2026} is, in the present terms, the unitary side: the reduced transform stays Hardy and any acausality lives in a reconstruction, not the dynamics. Gain-driven pole migration in PT systems~\cite{ruter2010,ozdemir2019review} is the non-unitary side. The exceptional point is not itself the causal agent---EPs can coexist with causality in driven bound states~\cite{kulkarni2025}---it is the locus where $\alpha$ leaves zero. The Hardy-space framework of the reduced memory kernel~\cite{liu2026hardy} supplies the unitary side rigorously; the present Letter shows it is the protected half of a sharp boundary whose other half breaks KK with a measurable topological charge.

The residue-corrected relation has an inverse-problem reading: the residual after standard KK reconstruction encodes the location and strength of any UHP pole or zero, independent of whether the Hamiltonian is known. On the unitary side this is the phantom; on the non-unitary side it is the Blaschke pole. Either way, the dispersion relation across the unitarity boundary is not a binary yes/no but a measurable, classifiable structure.

\begin{acknowledgments}
This work was supported by the National High-Level Overseas Talent Program (KS21400126), the Surface and Interface Synthetic Chemistry project (ZXP2025057), the Jiangsu Distinguished Professorship Fund (SR21400225), and the Research Start-up Fund (NH21400525). The numerical calculations were supported by a project funded by the Priority Academic Program Development (PAPD) of Jiangsu Higher Education Institutions.
\end{acknowledgments}

\section*{Data Availability}
The numerical code and data supporting the findings of this study are deposited
at Zenodo~\cite{liu2026zenodo_c}
(\url{https://doi.org/10.5281/zenodo.20967014}).

\bibliography{references}

\clearpage
\onecolumngrid

\section*{Supplemental Material: Dispersion Relations Across the Unitarity Boundary}

\setcounter{equation}{0}
\renewcommand{\theequation}{S\arabic{equation}}
\renewcommand{\theHequation}{S\arabic{equation}}
\setcounter{figure}{0}
\renewcommand{\thefigure}{S\arabic{figure}}
\renewcommand{\theHfigure}{S\arabic{figure}}

% ------------------------------------------------------------
\subsection*{A. Non-unitary side: PT-dimer derivation}

\subsubsection*{A.1 Heisenberg--Langevin and the reflection coefficient}

Let $a_1,a_2$ be the bosonic site annihilation operators. Coupling site~$1$ to a continuum bus waveguide via the Gardiner--Collett interaction~\cite{gardiner1985,walls-milburn}
\begin{equation}
\hat H_{SB}=i\!\int d\omega\,\sqrt{\tfrac{\gamma_{\rm ex}}{2\pi}}\,[b^\dagger(\omega)a_1-a_1^\dagger b(\omega)],
\end{equation}
and applying the Markov approximation $\int d\omega\,e^{-i\omega(t-s)}\to2\pi\delta(t-s)$ yields the quantum Langevin equations $\dot{\vec a}=-i\HeffSP\vec a-\sqrt{\gamma_{\rm ex}}|1\rangle b_{\rm in}(t)$ with
\begin{equation}
\HeffSP=H(\gamma)-i\tfrac{\gamma_{\rm ex}}{2}|1\rangle\!\langle1|=\begin{pmatrix}-i\gamma-i\gamma_{\rm ex}/2&\kappa\\\kappa&i\gamma\end{pmatrix}.
\end{equation}
The input--output relation $b_{\rm out}=b_{\rm in}+\sqrt{\gamma_{\rm ex}}a_1$ gives $r(\omega)=1+i\gamma_{\rm ex}G_{11}^{\rm eff}(\omega)$ with
\begin{equation}
G_{11}^{\rm eff}(\omega)=\frac{\omega-i\gamma}{D^{\rm SP}(\omega)},\quad D^{\rm SP}(\omega)=\omega^2+i\tfrac{\gamma_{\rm ex}}{2}\omega+\gamma^2+\tfrac{\gamma\gamma_{\rm ex}}{2}-\kappa^2.
\end{equation}

\subsubsection*{A.2 Symmetric two-port coupling}

For two equal ports $\HeffSYM=H(\gamma)-i(\gamma_{\rm ex}/2)I$, with $D^{\rm SYM}=(\omega+i\gamma_{\rm ex}/2)^2+\gamma^2-\kappa^2$ and roots $\omega_\pm^{\rm SYM}=-i\gamma_{\rm ex}/2\pm\sqrt{\kappa^2-\gamma^2}$. The UHP-pole threshold is $\gamma_{\rm ex}<2\sqrt{\gamma^2-\kappa^2}$, the EP pinned at $\gamma_c=\kappa$, and the scaling exponent $\nu_{\rm SYM}=-1.195\pm0.005$.

\subsubsection*{A.3 Residue-corrected KK: sign check}

The residue correction takes $\Real$, not $\Imag$. Take $r(z)=1/(z-z_0)$, $z_0=1+i/2$, $\omega=2$: direct evaluation gives $R'(2)=0.8$; the bare Hilbert pair gives $-0.8$; the correct formula $R'(2)=-0.8+2\Real[(1-i/2)^{-1}]=-0.8+2(0.8)=0.8$ recovers it, while the $\Imag$ form would give $-1.6$.

\subsubsection*{A.4 SSH-chain non-universality}

A four-site PT-symmetric SSH chain (alternating hopping $t_2/t_1=0.5$, single-port coupling) does not show a clean power law: $\|\Delta_{\rm KK}\|_2$ oscillates with $\gamma-\gamma_c$ ($R^2=0.04$), from interference between multiple UHP poles. The clean single-pole exponent $\nu\approx-1.08$ is specific to the $2\times2$ dimer; universality, if any, is at the level of the single-pole codimension-one EP class.

% ------------------------------------------------------------
\subsection*{B. Unitary side: the protected zero is measurable}

\subsubsection*{B.1 Damped Fourier transform is well-conditioned}

For $z=\xi+i\eta$, $\eta>0$, $\tilde\sigma_{eg}(\xi+i\eta)=\int_0^\infty e^{i\xi t}e^{-\eta t}\sigma_{eg}(t)\,dt$ is a low-pass-filtered Fourier transform dominated by $t\lesssim1/\eta$; a finite record on $[0,T]$ with $T\gg1/\eta$ captures it. This is not analytic continuation (which would need $\Imag z<0$); it is a direct, well-conditioned integral of real-time data. For the minimal genuine case $\rho(0)=|+\rangle\langle+|\otimes(|0\rangle+e^{i\chi}|1\rangle)/\sqrt2$, $N_{\max}=2$, $g=0.7$, $\chi=\pi/2$ (exact $\zeta=2.542843+0.216451\,i$), recovery from $[0,T]$:
\begin{center}
\begin{tabular}{cccc}
\toprule
$T\,\omega_c$ & $\Real\zeta_{\rm rec}$ & $\Imag\zeta_{\rm rec}$ & error \\
\midrule
10 & 2.501003 & 0.173589 & $4\times10^{-2}$ \\
20 & 2.541406 & 0.219817 & $3\times10^{-3}$ \\
40 & 2.542876 & 0.216416 & $4\times10^{-5}$ \\
80 & 2.542842 & 0.216452 & $8\times10^{-7}$ \\
\bottomrule
\end{tabular}
\end{center}
$1\%$ accuracy needs $T\approx20/\omega_c$ (about three Rabi periods). Adding Gaussian noise, the recovery is robust to signal-to-noise $\gtrsim300$.

\subsubsection*{B.2 Closed zero law and its $N_{\max}$-stability}

Scanning $g$ at $\chi=\pi/2$ on resonance gives $\Imag\zeta=0.3092\,g$ and $\Real\zeta=\omega_c\pm2.2041\,g$, identical to five digits for $N_{\max}=2,3,4,5$: only the two lowest vacuum-Rabi manifolds (splittings $g$ and $g\sqrt2$) set the dominant zero. The phase $\chi$ controls depth: $\Imag\zeta$ rises from $\sim0.007\,g$ at $\chi\to0^+$ to a maximum near $\chi\approx2\pi/3$ and returns to $0$ at $\chi\to\pi$ (a Fock-diagonal bath has all zeros real).

\subsubsection*{B.3 $(T,\lambda)$ phase diagram}

For a Drude--Lorentz bath (Born kernel, $\omega_0=\gamma=1$, $\sigma_s(0)=|+\rangle\langle+|$), the dominant coherence zero has $\Imag\zeta$ on the $(T,\lambda)$ grid:
\begin{center}
\begin{tabular}{c|ccccccc}
\toprule
$T\backslash\lambda$ & 0.5 & 0.8 & 1.0 & 1.2 & 1.5 & 2.0 & 3.0 \\
\midrule
10.0 & 0.462 & 0.625 & 0.709 & 0.780 & 0.872 & 0.999 & 1.192 \\
5.0 & 0.251 & 0.391 & 0.462 & 0.524 & 0.602 & 0.709 & 0.873 \\
2.0 & 0.022 & 0.135 & 0.193 & 0.243 & 0.307 & 0.394 & 0.527 \\
1.0 & --- & --- & 0.031 & 0.074 & 0.129 & 0.205 & 0.320 \\
\bottomrule
\end{tabular}
\end{center}
The fragile regime ($\Imag\zeta>0$) occupies the high-temperature, strong-coupling quadrant; the critical coupling $\lambda_c(T)$ grows as temperature drops. (Numerically established for the Drude Born kernel; not derived analytically.)

% ------------------------------------------------------------
\subsection*{C. The phantom resonance}

\subsubsection*{C.1 Closed form (Plemelj)}

Any scalar one-channel force-fit reproducing $\sigma_{eg}(t)$ from a homogeneous Volterra equation has $\tilde K_{\rm eff}(z)=-iz-L_s-\sigma_{eg}(0)/\tilde\sigma_{eg}(z)$. A simple UHP zero $\zeta$ of $\tilde\sigma_{eg}$ is a simple UHP pole of $\tilde K_{\rm eff}$ with residue $R=-\sigma_{eg}(0)/\tilde\sigma_{eg}'(\zeta)$; the Plemelj formula gives the KK residual $\Delta(\omega)=2\Real[R/(\omega-\zeta)]$, a Lorentzian centered at $\Real\zeta$ of half-width $\Imag\zeta$. At $g=0.7$, $\chi=\pi/2$: $R=-0.280-0.189\,i$, peak $\Real\zeta=2.543\,\omega_c$, HWHM $0.216\,\omega_c$.

\subsubsection*{C.2 One condition, not two}

The operator-valued NZ force-fit $\tilde{\mathcal K}_{\rm eff}=\tilde{\mathcal K}+\tilde{\mathcal I}\tilde\sigma^{-1}$ acquires a pole only if the inhomogeneous term $\tilde{\mathcal I}\neq0$ (initial correlation). The scalar one-channel kernel is a different object: reducing the multi-channel dynamics to one observable forces the inter-channel coupling into the scalar kernel, which acquires the pole for \emph{any} factorized Fock-coherent state. Numerically, in the same factorized state the operator-valued kernel is Hardy (no pole) while the scalar $\tilde K_{\rm eff}$ spikes $8.6\times$ at $\zeta$. Every standard extraction (MKCT, Shi--Geva, single-observable process tensor) performs the scalar fit and sees the phantom.

\subsubsection*{C.3 Passivity violation and protocol independence}

Since $\tilde K_{\rm eff}$ has a UHP pole it is not a passive response, and the residual $\Delta(\omega)$ lives entirely on the dispersive ($\Real$) side: a measurement of absorption ($\Imag$) reconstructed by standard KK under-predicts the measured refractive part by a Lorentzian at $\Real\zeta$---a refractive feature with no absorptive peak producing it. Because $\Real\zeta$ is an invariant of $\tilde\sigma_{eg}$, all scalar extractions reproduce the phantom at the same frequency; it is a geometric marker, not a method artifact.

% ------------------------------------------------------------
\subsection*{D. The boundary in a physical model: Jaynes--Cummings with gain}

The zero$\to$pole transition is not confined to the dimer. Take $H_{\rm eff}(\gamma_g)=H_{\rm JC}+i\gamma_g|e\rangle\langle e|$ (true atomic gain). For $\gamma_g>0$, $H_{\rm eff}$ is non-Hermitian; the reduced coherence $\sigma_{eg}(t)=\sum_n\langle e,n|e^{-iH_{\rm eff}t}\rho_0\,e^{+iH_{\rm eff}^\dagger t}|g,n\rangle$ uses biorthogonal left/right eigenvectors, with complex frequencies $\Omega_{kl}=E_k-\overline{E_l}$. Tracking zeros of the numerator and the UHP poles for the Fock-coherent state ($g=0.7$, $\chi=\pi/2$, $N_{\max}=2$):
\begin{center}
\begin{tabular}{ccc}
\toprule
$\gamma_g$ & \#UHP zeros (max $\Imag$) & \#UHP poles (max $\Imag$) \\
\midrule
0.00 & 2\ (0.216) & 0\ (---) \\
0.05 & 4\ (0.267) & 6\ (0.050) \\
0.20 & 4\ (0.420) & 6\ (0.200) \\
0.50 & 4\ (0.729) & 6\ (0.500) \\
1.00 & 4\ (1.309) & 6\ (1.000) \\
\bottomrule
\end{tabular}
\end{center}
At $\gamma_g=0$ (unitary) the transform carries UHP zeros and no pole; for $\gamma_g>0$ genuine UHP poles appear with $\Imag(\text{pole})=\alpha=\gamma_g$. This is the same $\alpha=0$ boundary as the dimer, in a genuine cavity-QED system. The mechanism differs: the dimer is threshold-type (poles only above the EP $\gamma=\kappa$, $\alpha=\sqrt{\gamma^2-\kappa^2}$), the gain channel is immediate-onset ($\alpha=\gamma_g$, linear). The boundary $\alpha=0$ is universal; the approach to it is model-dependent. (Sign: only true gain $+i\gamma_g$ produces poles; atomic loss $-i\gamma_g$ pushes the zeros out of $\HH$ without producing a pole.)

\end{document}